\documentclass[conference]{IEEEtran}
\IEEEoverridecommandlockouts
\usepackage{cite}
\usepackage{amsmath,amssymb,amsfonts}
\usepackage{algorithmic}
\usepackage{graphicx}
\usepackage{textcomp}
\usepackage{xcolor}
\usepackage{multirow}
\usepackage{diagbox}
\usepackage{makecell}
\usepackage{subfigure}
\def\BibTeX{{\rm B\kern-.05em{\sc i\kern-.025em b}\kern-.08em
    T\kern-.1667em\lower.7ex\hbox{E}\kern-.125emX}}
\begin{document}

\title{RAIL: An Accurate and Fast Angle-inferred Localization Algorithm for UAV-WSN Systems\\}

\author{
Ze Zhang, Qian Dong*\\
School of Advanced Technology, Xi'an Jiaotong-Liverpool University, Suzhou, China \\
Emails: Ze.Zhang20@alumni.xjtlu.edu.cn, Qian.Dong@xjtlu.edu.cn
}
\maketitle

\begin{abstract}
Location information is a fundamental requirement for unmanned aerial vehicles (UAVs) and other wireless sensor networks (WSNs). However, accurately and efficiently localizing sensor nodes with diverse functionalities remains a significant challenge, particularly in a hardware-constrained environment. To address this issue and enhance the applicability of artificial intelligence (AI), this paper proposes a localization algorithm that does not require additional hardware. Specifically, the angle between a node and the anchor nodes is estimated based on the received signal strength indication (RSSI). A subsequent localization strategy leverages the inferred angular relationships in conjunction with a bounding box. Experimental evaluations in three scenarios with varying number of nodes demonstrate that the proposed method achieves substantial improvements in localization accuracy, reducing the average error by 72.4\% compared to the Min-Max and RSSI-based DV-Hop algorithms, respectively.
\end{abstract}


\section{Introduction}
As the terrestrial economy of human society becomes increasingly saturated, the low-altitude economy (LAE) within the context of the Internet of Things (IoTs) has emerged as a promising area for research and development. In LAE, unmanned aerial vehicles (UAVs) represent the core component at low altitudes. Their high flexibility, cost-effectiveness, and multi-functionality enable them to efficiently perform a wide range of tasks\cite{b1}. UAVs play an indispensable role in scenarios such as environmental monitoring in rugged or narrow terrain\cite{b2}\cite{b3}, disaster response\cite{b4}\cite{b5}, agricultural irrigation and production\cite{b6}\cite{b7}, and military operations\cite{b8}\cite{b9}. As UAV applications continue to expand, the complexity and dynamism of their tasks demand increasing levels of cooperation among multiple UAVs\cite{b10}. In such collaborative UAV systems, accurate localization of each UAV is critical, directly affecting the efficiency and safety of group operations.

Currently, the most commonly adopted UAV localization methods are based on visual input processed by artificial intelligence (AI). These approaches analyze continuous image frames using convolutional neural networks (CNNs)\cite{b11}, transformer architectures\cite{b12}, or depth estimation networks within simultaneous localization and mapping (SLAM) systems\cite{b13}. However, these methods are highly sensitive to environmental conditions,  such as lighting and visual obstructions, making them prone to failure, an issue that becomes particularly problematic in high-speed UAV swarms. Furthermore, due to hardware constraints, most UAVs are equipped only with fixed-angle cameras or cameras mounted on limited gimbals. As such, it is practical to determine the approximate directional relationship between UAVs beforehand and subsequently use vision systems for fine-grained localization. Although some advanced UAVs can utilize point cloud LiDAR for distance measurement, the high cost of such equipment significantly limits its widespread adoption.

To address these challenges, this study proposes a novel localization technique based on received signal strength indication (RSSI), which requires no additional hardware, thus maximizing the algorithm's applicability within UAV networks. The detailed flow chart of the proposed RSSI angle-inferred localization (RAIL) algorithm is presented in Fig.~\ref{fig1}. The key contributions of this study are as follows:

\begin{figure}[htbp]
\centerline{\includegraphics[width=2.8in]{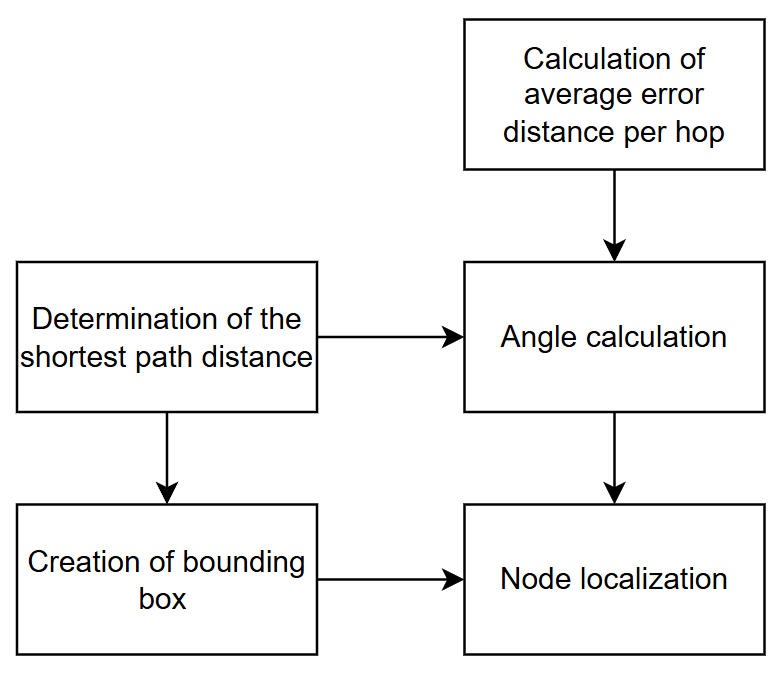}}
\caption{Main flowchart of the RAIL algorithm}
\label{fig1}
\end{figure}

\begin{enumerate}
    \item \textbf{RSSI-based angle inference between nodes.} By leveraging the multi-hop relationships between nodes and anchor nodes, the proposed method infers the angle of the line connecting any node with an anchor node. This enables the estimation of the angle between any two nodes.
    \item \textbf{Accurate and fast node localization.} The method combines directional relationships with bounding box constraints to localize nodes efficiently and robustly, offering high generalizability.
    \item \textbf{Practical value for AI models.} The RAIL algorithm can extract implicit spatial information, such as direction and relative position, purely from RSSI data. This improves the effectiveness and robustness of AI model training and is also valuable for data generation and baseline evaluation tasks.
\end{enumerate}

The rest of this paper is structured as follows: Section \uppercase\expandafter{\romannumeral2} reviews the technical background and related work in this field and introduces the technologies used. Section \uppercase\expandafter{\romannumeral3} details the complete RAIL algorithm. Section \uppercase\expandafter{\romannumeral4} evaluates its localization performance and analyzes the underlying causes of the observed outcomes. Section \uppercase\expandafter{\romannumeral5} presents the conclusion of the paper.

\section{Technical Background}
This section provides an overview of the main distance estimation techniques and localization algorithms commonly used in wireless sensor networks (WSNs). The distance estimation method adopted in this study is described in detail, followed by a detailed introduction to two widely used localization algorithms: the Min-Max algorithm and the RSSI-based DV-hop algorithm. These two algorithms are selected as benchmarks for comparison with the proposed method, and their performance will be evaluated along with our algorithm in the numerical results section.

\subsection{Distance Estimation}
Node localization methods in WSNs are generally classified into two main categories: range-free and range-based approaches\cite{b14}. The fundamental distinction between them lies in whether accurate distance measurements within one-hop communication range are required.

Range-free localization methods rely solely on hop count information between nodes to estimate positions. These methods do not require additional hardware, offering advantages in terms of lower algorithmic complexity and reduced energy consumption. However, their main drawback is their limited localization accuracy, making them less suitable for scenarios that require high precision.

In contrast, range-based localization techniques are commonly employed when higher localization accuracy is required. These methods utilize explicit distance measurements between nodes, typically through techniques such as RSSI\cite{b15}, Time of Arrival (TOA)\cite{b16}, and Time Difference of Arrival (TDOA)\cite{b17}. In addition, there is a method called Angle of Arrival (AoA)\cite{b18}, which estimates angular relationships. Each method has its own advantages and is suitable for different application scenarios. Notably, RSSI operates without the need for extra hardware, which makes it particularly economical in large-scale WSNs deployments.

This paper proposes a range-based localization algorithm that is adaptable to any distance-based ranging approach. However, considering the cost effectiveness of RSSI and its extensive use in recent research, this work adopts RSSI as the primary ranging method. Signal transmission between two nodes experiences attenuation in any propagation medium. Distance estimation between nodes in the RSSI-based method is achieved by measuring how much the signal weakens. The wireless signal path loss model used for RSSI-based distance estimation is given as follows:
\begin{equation}
RSSI(d)= RSSI(d_0)-10n\log_{10}(\frac{d}{d_0})+N(0,\sigma^2)
\label{eq1}\end{equation}

In \eqref{eq1}, the parameter $RSSI(d)$ denotes the received signal strength at a distance \textit{d}, while $RSSI(d_0)$ represents the received signal strength at a reference distance $d_0$ (typically set at 1 meter). The path loss exponent $n$ varies depending to the characteristics of the propagation environment. The term $N(0,\sigma^2)$ accounts for the random noise in the system. The final estimated distance $d$ between the two nodes is obtained by solving the path loss model equation. 

\subsection{Localization Algorithm}
Min-Max, DV-hop, APIT, and Amorphous are representative and mature range-free localization algorithms. However, as the demand for higher localization accuracy increases, RSSI-based ranging has been incorporated into these methods to improve performance.

The original Min-Max algorithm estimates the node's location by constructing a rectangular region based on the product of the number of hops from the unknown node to each anchor node and the maximum communication range. The final estimated coordinates are determined by the intersection point of the diagonals of the overlapping rectangles derived from multiple anchor nodes. Although this approach is computationally simple and highly versatile, it suffers from significant localization errors when all anchor nodes are located on the same side of the unknown node, a condition frequently encountered in practice. In \cite{b19} and \cite{b20}, RSSI is introduced into the Min-Max algorithm to refine the size of the bounding rectangle and improve localization accuracy. However, its performance remains highly sensitive to the spatial configuration between the anchor and unknown nodes.

The DV-hop algorithm begins by estimating the average hop distance across the network using the actual distances and hop counts between the anchor nodes. The distance between an anchor node and an unknown node is then approximated by multiplying the number of hops by the average hop distance. Finally, the node's coordinates are estimated via a least squares fitting approach based on multiple such distances. However, in environments with uneven node density, the resulting localization errors can be substantial. Directly using RSSI multi-hop accumulation distance measurement instead of average hop distance can reduce this effect. In \cite{b21}, RSSI was introduced as a correction factor, and an improved coordinate estimation method replaced the traditional least squares algorithm. Although this modification significantly enhanced localization accuracy, it also introduced a higher computational burden.

\section{Methodology}
For clarity, this section illustrates the proposed method within a two-dimensional WSN region. This method can be readily extended to three-dimensional space. To validate the general applicability of the system, it is assumed that the network includes only three anchor nodes with predetermined coordinates, representing the minimal requirement. All other unknown nodes are randomly deployed throughout the region. However, isolated nodes are not allowed, meaning each node must have at least one neighboring node to ensure network connectivity. This algorithm is well-suited to the distributed nature of WSNs. This method is capable of efficiently locating all unknown nodes with ease. However, for clarity and simplicity, this section focuses on the localization of a single unknown target node $T$. All remaining parameters referenced in this section are formally defined in TABLE~\ref{tab1}.

\begin{table}
\caption{Parameter List}
\label{table}
\setlength{\tabcolsep}{5pt}
\begin{tabular}{|p{45pt}|p{185pt}|}
\hline
Parameter& Definition\\
\hline
$A_1, A_2, A_3$& The anchor nodes with node identifiers $1$, $2$, and $3$\\
$D_{T,A_i}^{g}$& The $g^{th}$ distance between $T$ and $A_i$\\
$N_n$& The node with node identifiers $n$\\
$d_{N_{n-1},N_n}$& The distance between the node $N_{n-1}$ and $N_n$\\
$SD_{T,A_i}^{g}$& The shortest $g^{th}$ distance between $T$ and $A_i$\\
$x_i,y_i$& The two-dimensional coordinates of $A_i$\\
$x_T,y_T$& The two-dimensional coordinates of $T$\\
$e$& The average error per hop in the system\\
$TD_{A_1,A_2,A_3}$& The sum of the true distances between $A_1$, $A_2$, and $A_3$\\
$HOP_{A_1,A_2}$& The minimum number of hops along the shortest path between $A_1$ and $A_2$\\
$\angle A_2A_1T$& The angle between the line $A_2A_1$ and $A_1T$ calculated by the algorithm\\
\hline
\end{tabular}
\label{tab1}
\end{table}

\subsection{Creation of Bounding Box}
By utilizing the multi-hop communication mechanism commonly employed in WSNs, two nodes separated by a distance greater than the maximum single-hop communication range can still exchange information through intermediate nodes. This method provides benefits in both communication performance and energy consumption. Neighboring nodes can use RSSI to directly measure the distance between two nodes. In this context, the distance from $A_1$, $A_2$, and $A_3$ to an unknown target node $T$ is calculated as the sum of the distances over each hop, and is expressed as: 
\begin{equation}
D_{T,A_i}^{g}= d_{T,N_1} + d_{N_1,N_2}+ d_{N_2,N_3}+\dots+ d_{N_{n-1},N_n}+d_{N_n,A_i}
\label{eq2}\end{equation}

In \eqref{eq2}, $D_{T,A_i}^{g}$ denotes the total estimated distance obtained by multi-hop communication from $A_i$ to unknown node $T$, where $i = 1, 2, 3$. Since multiple paths may exist between $A_i$ and unknown node $T$, the parameter $g$ represents the total number of such paths. The parameter $N_i$ represents the relay node on the path from $A_i$ to $T$, and $n$ represents the number of relay nodes. The value of $d_{N_1,N_2}$ for a one-hop connection can be directly estimated using RSSI measurements. According to the equation, the total number of hops in a given path is $n+1$. It is worth noting that in scenarios where more than three anchor nodes are available, selecting the three anchor nodes closest to the target node can enhance the localization accuracy while conserving the limited energy resources of the nodes. The shortest path distance from $A_i$ to unknown node $T$ can be expressed as:
\begin{equation}
SD_{T,A_i}^{g}=
\begin{cases}
D_{T,A_i}^{g} & D_{T,A_i}^{g}<SD_{T,A_i}^{g-1}\\
SD_{T,A_i}^{g-1} & D_{T,A_i}^{g} \geq SD_{T,A_i}^{g-1}\\
\end{cases}
\label{eq3}\end{equation}

In \eqref{eq3}, $SD_{T,A_i}^{g}$ represents the shortest path distance from $A_i$ to the unknown node $T$ in $g$ iterations. It is worth noting that the parameter $g$ in this context differs from that in \eqref{eq2}. Here, $g$ denotes the number of iterations, ranging from 2 to a predefined maximum value $g_{max}$. Consequently, $SD_{T,A_i}^{g_{max}}$ in \eqref{eq3} specifically refers to the length of the shortest path from three anchor nodes $A_i$ to the unknown node $T$. Although the shortest multi-hop distance between an anchor node and an unknown node can be estimated, a discrepancy still exists between this calculated value and the actual distance. This is primarily due to the inherent nature of multi-hop communication, where it is unlikely that all intermediate relay nodes lie precisely along the straight line connecting the anchor node and the unknown node. Therefore, the calculated shortest path tends to be curved or deviated, leading to a value of $SD_{T,A_i}^{g_{max}}$ that is slightly greater than the true physical distance. 

At this stage, the positions of the three anchor nodes and their shortest distances to the unknown node have been determined. However, due to the absence of directional information between each anchor node and the unknown node $T$, the exact location of $T$ cannot be determined accurately. However, for the three anchor nodes $A_i$, with known coordinates $(x_i,y_i)$, the coordinates of $T$ must satisfy the following equation: 
\begin{equation}
\begin{cases}
x_i - SD_{T,A_i}^{g_{max}} < x_{T} < x_i + SD_{T,A_i}^{g_{max}}\\
y_i - SD_{T,A_i}^{g_{max}} < y_{T} < y_i + SD_{T,A_i}^{g_{max}}\\
\end{cases}
\label{eq4}\end{equation}

\begin{figure}[htbp]
\centerline{\includegraphics[width=3.2in]{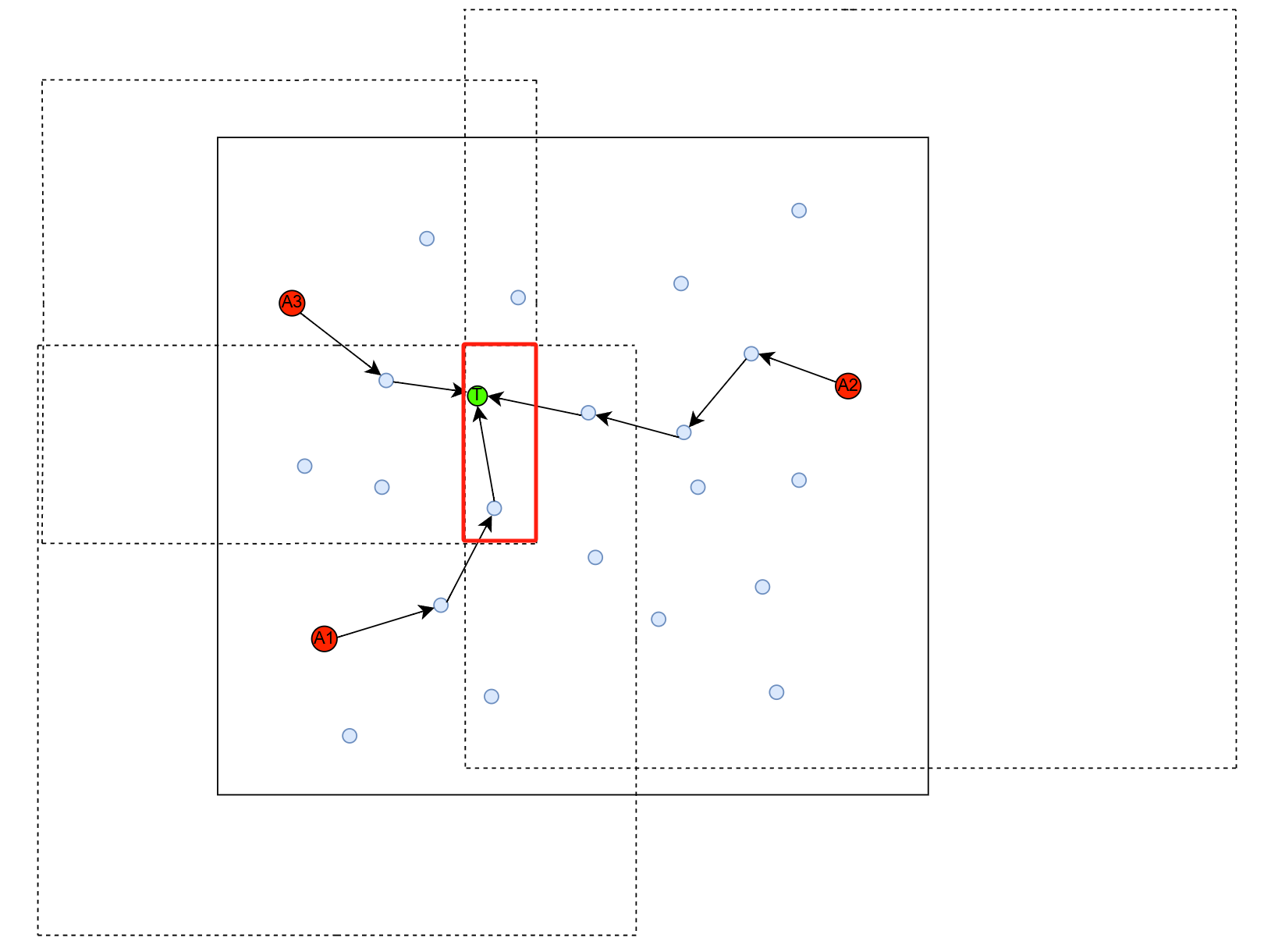}}
\caption{The bounding box of $T$ is determined based on the positions and estimated ranges of the three anchor nodes. Each anchor node’s estimated range, calculated using \eqref{eq4}, is represented by a black dashed rectangle. The intersection of these three individual rectangles forms the intersectional bounding box within which the unknown node $T$ is constrained. This bounding box is depicted as a red solid rectangle in the diagram.}
\label{fig2}
\end{figure}

Using \eqref{eq4}, each of the three anchor nodes can independently estimate an interval for both the horizontal and vertical coordinates of $T$. The intersection of these three intervals in both dimensions forms the bounding box within which the unknown node is constrained. As previously discussed, due to the bending nature of multi-hop communication paths, the estimated distance from an anchor node to the unknown node often exceeds the actual distance. Consequently, as the spatial separation between the anchor nodes and the unknown node increases, the resulting bounding box may become excessively large. Therefore, further precision is critical for the location of $T$. However, it is important to emphasize that, assuming an error-free distance estimate, the true location of $T$ will always lie within the calculated bounding box, regardless of the scenario. A specific example is shown in Fig.~\ref{fig2}.

\subsection{Angle Prediction}
In this step, the RSSI-based distance estimates are utilized to approximate the angle between each anchor node and the unknown node. However, as previously discussed, the bending of the shortest multi-hop path causes the estimated distance between them to be greater than the actual distance. To address this discrepancy, it is necessary to calculate the average error introduced per hop and apply a correction to the estimated distance. This correction aims to compensate for the cumulative deviation and minimize the overall localization error. The average error per hop in the system can be expressed as:
\begin{equation}
e= \frac{SD_{A_1,A_2}^{g_{max}}+SD_{A_1,A_3}^{g_{max}}+SD_{A_2,A_3}^{g_{max}}-TD_{A_1,A_2,A_3}}{HOP_{A_1,A_2}+HOP_{A_1,A_3}+HOP_{A_2,A_3}}
\label{eq5}\end{equation}

In \eqref{eq5}, $SD_{A_1,A_2}^{g_{max}}$ represents the shortest multi-hop path distance between $A_1$ and $A_2$, while $TD_{A_1,A_2,A_3}$ represents the sum of the true distances between $A_1$, $A_2$, and $A_3$. The parameter $HOP_{A_1,A_2}$ denotes the minimal number of hops along the shortest path between $A_1$ and $A_2$.

Although the path between an anchor node and the unknown node deviates from a straight line due to the nature of multi-hop routing and randomly distribution of nodes, its general direction remains approximately consistent. To ensure that the proposed method maintains low power consumption, high robustness, and improved accuracy, the direction of the unknown node is estimated using geometric information derived from three anchor nodes. Since the average error per hop of the system is calculated from the anchor nodes, the nodes located close to the anchor nodes are more suitable for this correction mechanism, thereby reducing the localization error.

\begin{figure}[htbp]
\centerline{\includegraphics[width=2.5in]{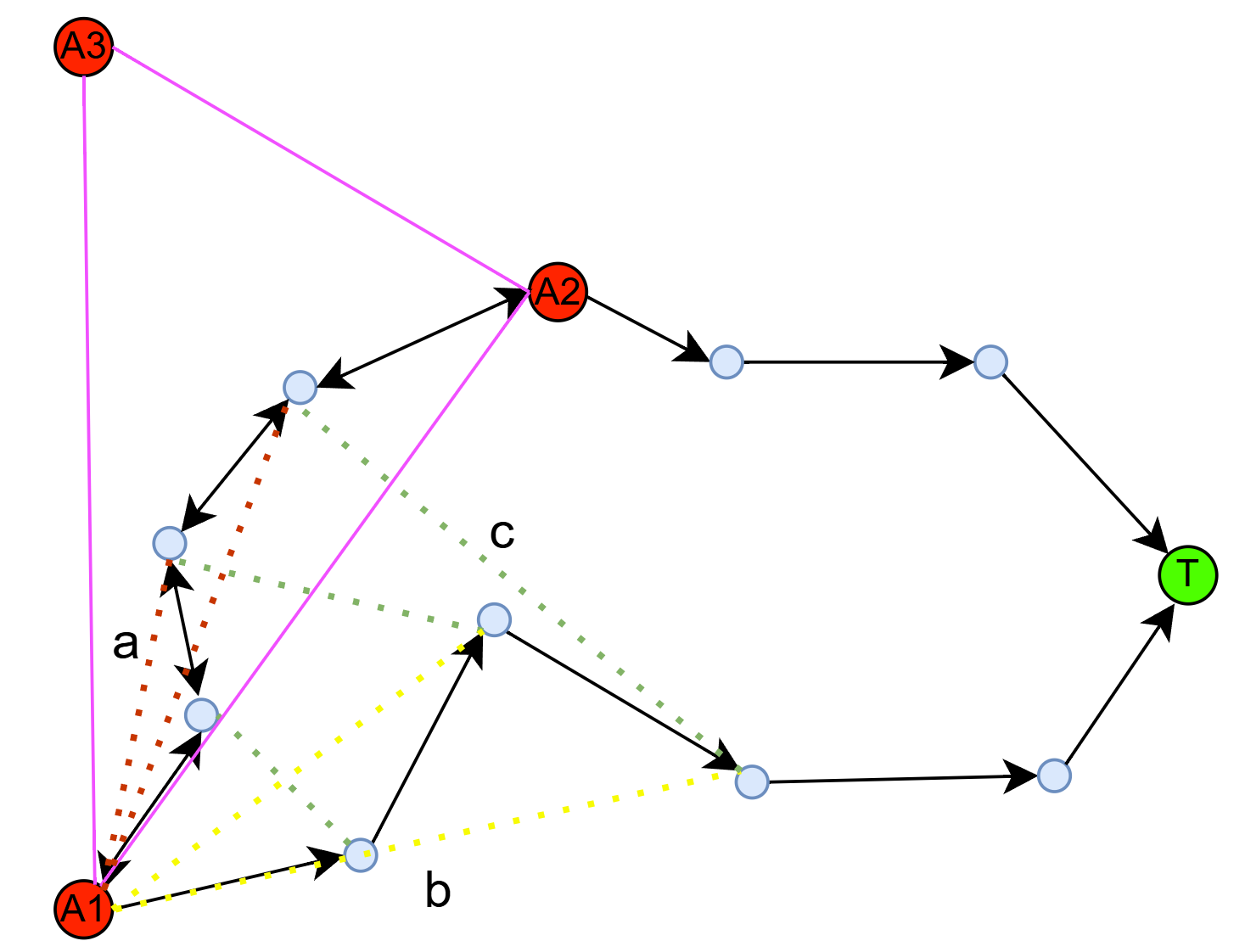}}
\caption{The segment formed by the first three nodes along the shortest path from $A_1$ to $A_2$ is defined as $a$ edge, and is depicted as a red line. Similarly, the segment formed by the first three nodes along the shortest path from $A_1$ to $T$ is defined as $b$ edge, represented by a yellow line. The line connecting the nodes on $a$ edge and $b$ edge with the same number of hops to $A_1$ is defined as $c$ edge.
}
\label{fig3}
\end{figure}

Referring to Fig.~\ref{fig3}, we define the line segment between anchor nodes $A_1$ and $A_2$ as the $a$ edge, the line segment between $A_1$ and the unknown node $T$ as the $b$ edge. For each hop index, the connection of nodes on $a$ edge and $b$ edge,  the nodes located at the same hop distance from $A_1$ via different routes to $A_2$ and $T$, is defined as $c$ edge. When the number of hops along any edge is greater than or equal to 2, a correction factor is applied by subtracting the error occurring on each hop $e$ from the measured edge length. Under this setup, the angle $\angle A_2A_1T$ can be computed as:
\begin{equation}
\angle A_2A_1T= cos^{-1}\frac{(a-e\cdot n)^2+(b-e\cdot n)^2-(c-e\cdot n)^2}{2(a-e\cdot n)(b-e\cdot n)}
\label{eq6}\end{equation}
 
The average of the three angles calculated using \eqref{eq6} is taken as the estimated angle between the line segment connecting anchor nodes $A_1$ and $A_2$, and the line segment connecting $A_1$ to the unknown node $T$. While the direction and position of the line $A_1A_2$ are known, it remains ambiguous whether the direction of $A_1T$ is clockwise or counterclockwise relative to $A_1A_2$. To resolve this uncertainty, the same method is applied to calculate the angle $\angle A_3A_1T$. The angle $\angle A_3A_1A_2$ is also known, which provides additional directional context and enables disambiguation of the orientation of the line segment $A_1T$.

\begin{figure}[htbp]
\centerline{\includegraphics[width=3.2in]{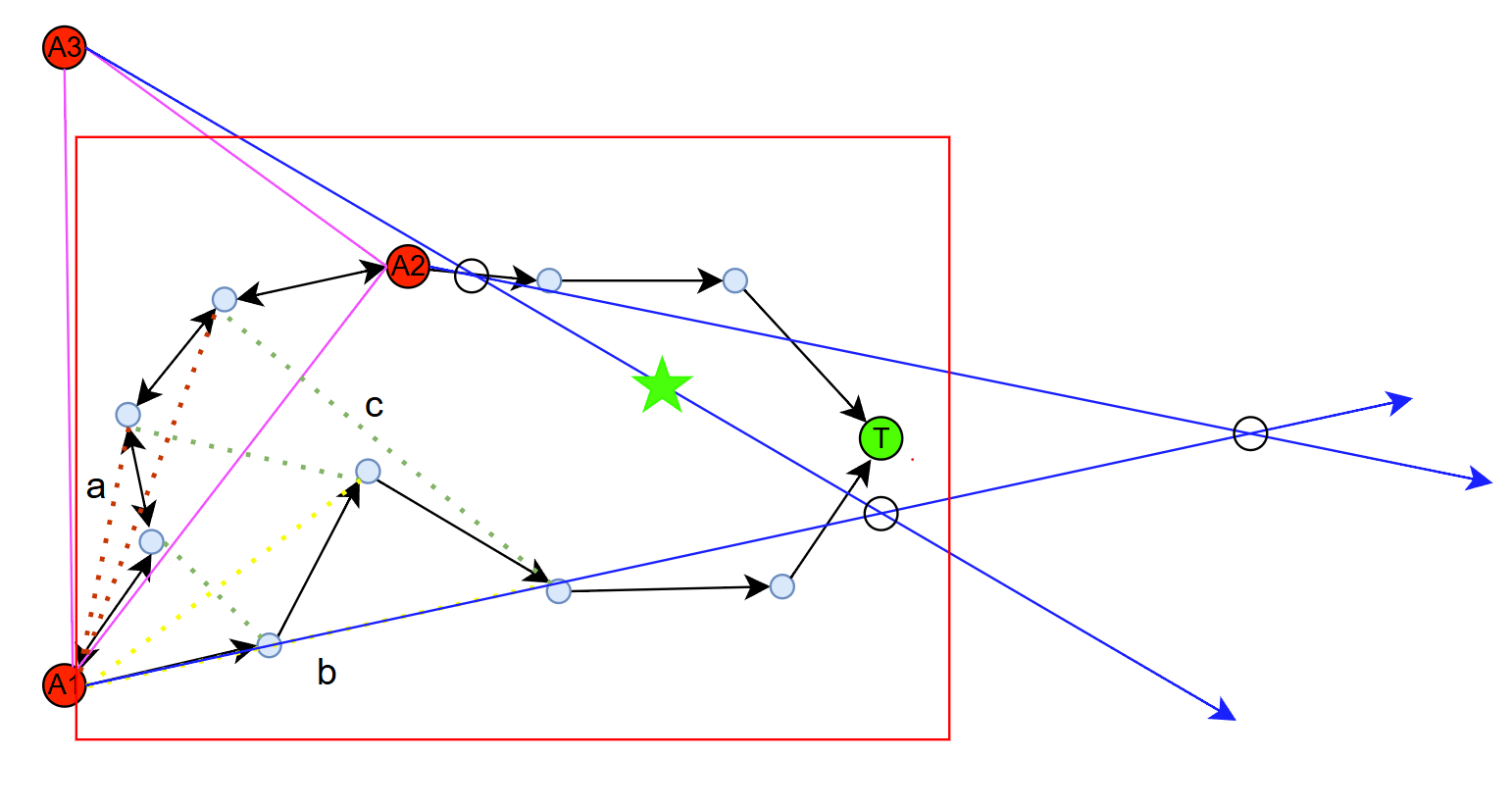}}
\caption{The red rectangle represents the bounding box of the unknown node $T$. The three directional rays, calculated based on the angular estimations from the anchor nodes, are depicted as blue lines with arrows indicating their directions. The intersection points of the rays are marked by black circles, while the final estimated location of $T$ is denoted by a green star. }
\label{fig4}
\end{figure}

Repetition of the above process can obtain three directional rays originating from the anchor nodes and pointing toward the estimated location of the unknown node $T$. These rays form the basis for localizing the unknown node and are illustrated in Fig.~\ref{fig4}.

\subsection{Precise Location}
Based on the analysis in the previous steps, the following conclusions can be drawn:

\begin{enumerate}
    \item \textbf{A bounding box} enclosing the unknown node $T$ can be established, ensuring that the estimated location of node lies within this confined region.
    \item \textbf{Three directional rays}, each originating from an anchor node and pointing toward the estimated location of $T$, are obtained.
\end{enumerate}

The final estimated position of the unknown node is obtained by analyzing the intersections of these rays with the bounding box. Four possible scenarios are considered:

\begin{enumerate}
    \item \textbf{Two or three ray intersections fall within (or on) the bounding box}:
In this case, the average of the horizontal and vertical coordinates of multiple intersections is calculated as the coordinate of $T$. If there are two intersections, the midpoint of the line segment connecting the two points is taken as the coordinate of node $T$. If there are three intersections, the centroid of the triangle formed by the intersections is used as the estimated coordinate of node $T$.
    \item \textbf{Only one ray intersection lies within (or on) the bounding box}:
In this case, the single intersection point is directly used as the coordinate of the unknown node $T$.
    \item \textbf{All ray intersections lie outside the bounding box}:
The intersection point closest to the bounding box is identified and projected onto the nearest edge of the bounding box using the shortest distance. The projected point on the bounding box is then taken as the estimated location of node $T$.
    \item \textbf{No intersections exist, either inside or outside the bounding box}:
In this rare case, the intersection point of the diagonals of the bounding box serves as the estimated coordinate of $T$.

\end{enumerate}
 
A specific example is illustrated in Fig.~\ref{fig4}. In this figure, three ray intersections are observed: two of which lie within the bounding box, while the third falls outside of it. This scenario corresponds to the first localization condition described previously. Consequently, the midpoint of the line segment connecting the two intersection points within the bounding box is used as the estimated coordinate of $T$.

\section{Numerical Results}
This section evaluates the localization accuracy of the RAIL algorithm under three different number of nodes scenarios. To assess its effectiveness, RAIL is compared with the Min-Max and the RSSI-based DV-hop localization algorithm using multiple datasets.

\subsection{Simulation Setups and Assessment Criteria}
The experiments were conducted in a $50 \times 50 m^2$ area, evaluating three scenarios with different node quantities: 100, 200, and 500. All nodes were randomly deployed within the region. To assess the extreme performance of each algorithm, only three anchor nodes were used, and they were not within one-hop communication range of each other. This configuration ensures that anchor nodes are not too close, preventing redundancy in multi-hop communication and reflects realistic deployment scenarios. The maximum communication radius for each node was set to $10m$. A smaller communication radius helps improve the precision of RSSI-based distance measurements. Accordingly, the standard deviation $\sigma$ of the Gaussian noise in \eqref{eq1} is assumed to be zero. The number of simulation for each algorithm under three node densities is 50. The specific simulation parameter settings are listed in TABLE~\ref{tab2}.

\begin{table}[htbp]
\caption{Simulation Parameters}
\label{table}
\setlength{\tabcolsep}{10pt}
\begin{tabular}{|p{95pt}|p{115pt}|}
\hline
Parameter& Parameter Values\\
\hline
Area of sensor nodes& $50 \times 50 m^2$\\
Number of unknown nodes& $100$, $200$, and $500$\\
Number of anchor nodes& $3$\\
Communication range& $10m$\\
Variance of noise& $0$\\
Number of simulation for each node density for each algorithm& $50$\\
\hline
\end{tabular}
\label{tab2}
\end{table}

To evaluate localization accuracy, the Euclidean distance between the estimated coordinates and the actual coordinates of $T$ is used as the performance metric. The calculation is defined as:
\begin{equation}
Error = \sqrt{(x_T^{true} - x_T^{est})^2 + (y_T^{true} - y_T^{est})^2}
\label{eq7}\end{equation}

The parameter $Error$ represents the Euclidean distance between the actual and estimated coordinates of $T$ in a single simulation. Specifically, $x_T^{true}$ and $y_T^{true}$ denote the true coordinates of the unknown node, while $x_T^{est}$ and $y_T^{est}$ represent the estimated coordinates obtained through localization. To facilitate a comprehensive comparison of localization performance, both the mean and standard deviation of the localization error are calculated for each algorithm across all simulations. These statistical metrics help evaluate not only the accuracy but also the stability of the localization methods under different node densities.

\subsection{Evaluation Results of  RAIL}
\begin{figure}[htbp]
\centering
\subfigure[Configuration with 100 unknown nodes]
{\includegraphics[width=3.5in]{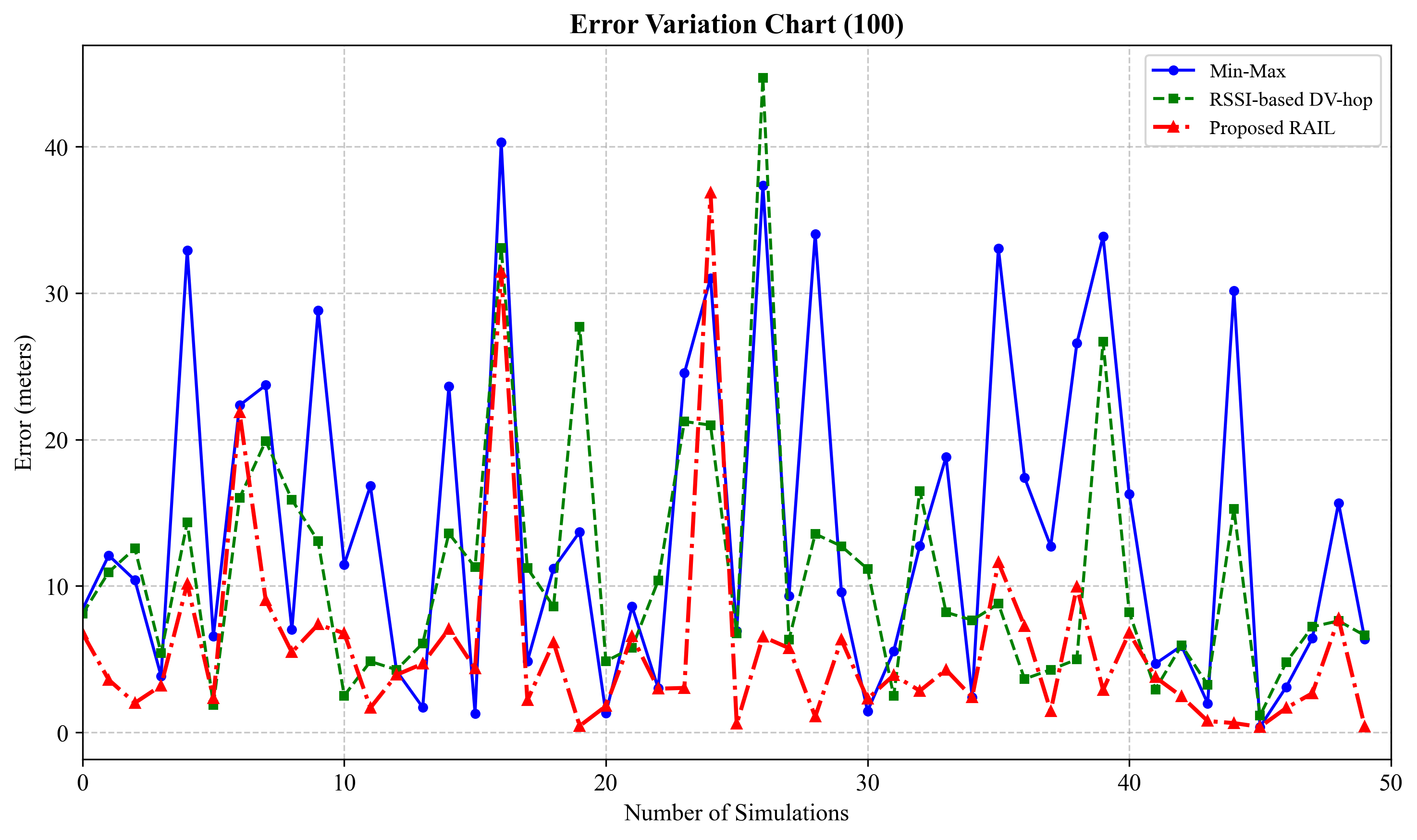}}
\subfigure[Configuration with 200 unknown nodes]
{\includegraphics[width=3.5in]{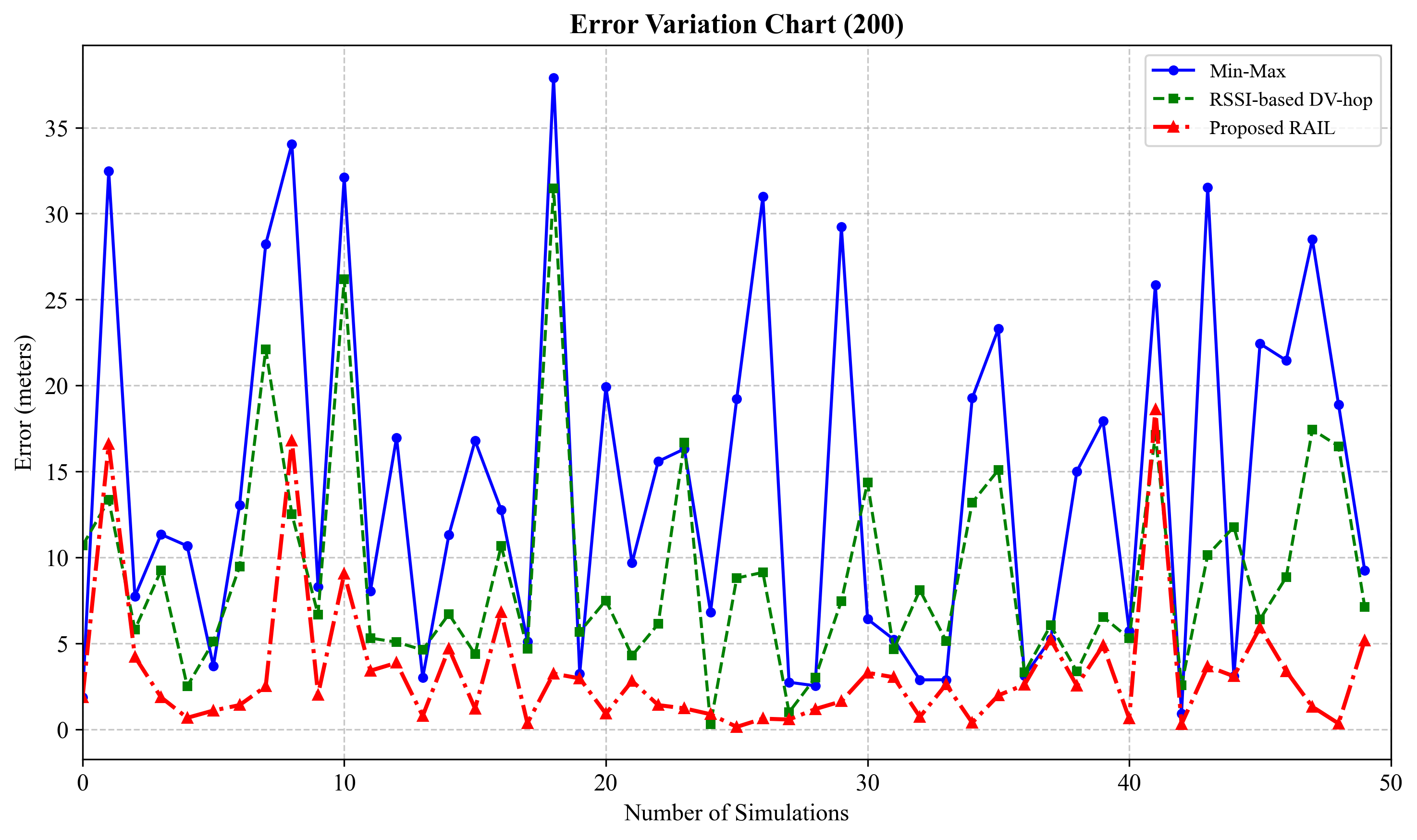}}
\subfigure[Configuration with 500 unknown nodes]
{\includegraphics[width=3.5in]{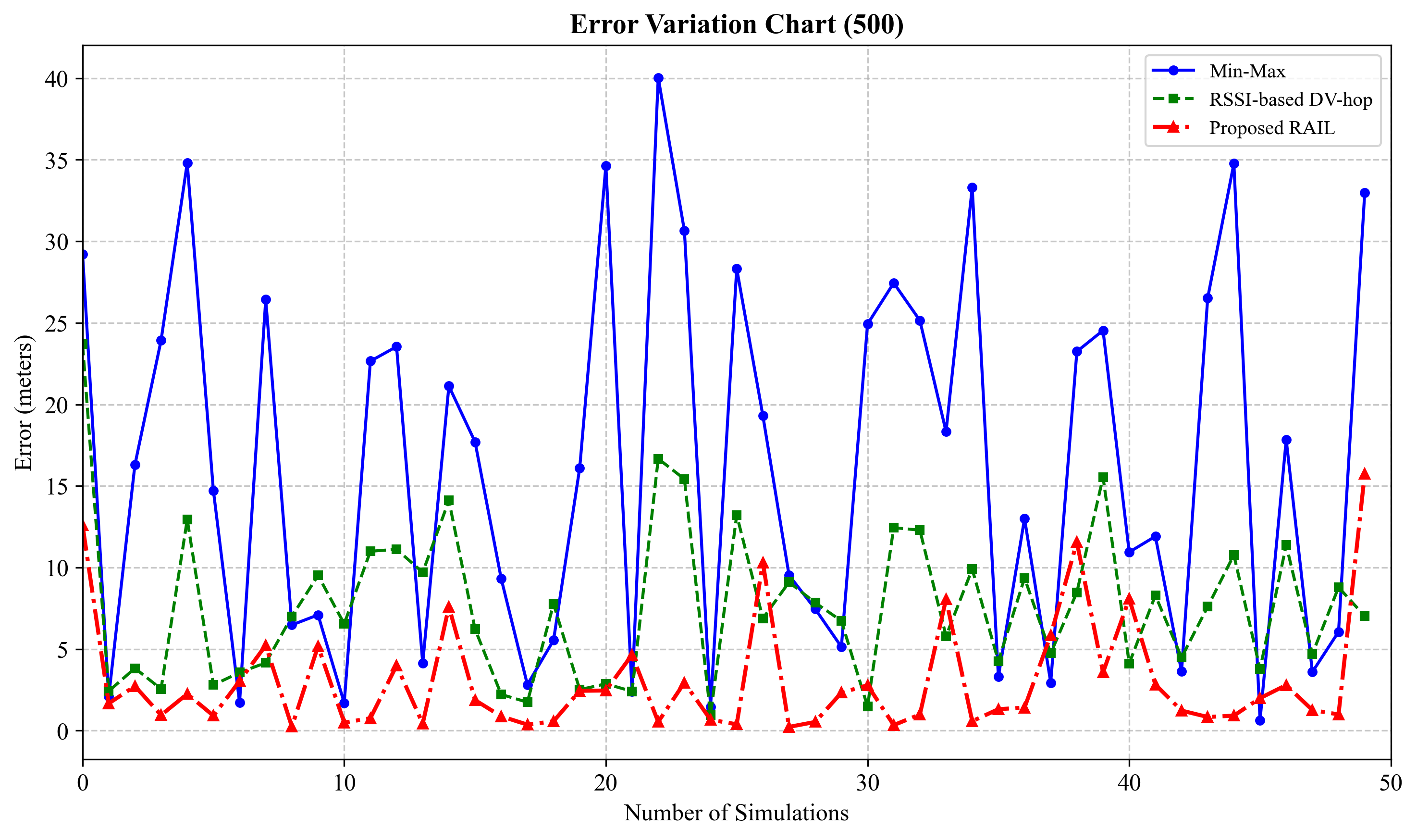}}
\caption{The localization error variation}
\label{fig5}
\end{figure}

According to the previously introduced error evaluation criteria for unknown nodes, Fig.~\ref{fig5} illustrates the error variation chart of the three localization methods under different number of nodes. Correspondingly, TABLE~\ref{tab3} presents the mean localization error and standard deviation error for the three methods across varying node densities.

\begin{table}[htbp]
\caption{The mean and standard deviation error of the three algorithms}
\begin{center}
\begin{tabular}{|>{\centering\arraybackslash}p{1.7cm}|>{\centering\arraybackslash}p{1cm}|c|>{\centering\arraybackslash}p{1.2cm}|>{\centering\arraybackslash}p{1.3cm}|}
\hline
  & \textbf{Number of Nodes} & \multirow{3}{*}{\textbf{Min-Max}} & \textbf{RSSI-based DV-hop} & \multirow{3}{*}{\textbf{\makecell{RAIL}}} \\
\hline
\multirow{3}{*}{\textbf{Mean Error}} & 100 & 14.1354 & 10.9239 & 5.7667 \\
                         & 200 & 14.4104 & 8.9895 & 3.3298 \\
                         & 500 & 16.2267 & 7.6543 & 3.0470\\
\hline
\multirow{3}{*}{\textbf{\makecell{Standard\\Deviation\\Error}}} 
                        & 100 & 11.3817 & 8.5082 & 6.9649 \\
                         & 200 & 10.4720 & 6.2584 & 4.0369 \\
                         & 500 & 11.5209 & 4.7224 & 3.5155 \\
\hline
\end{tabular}
\end{center}
\label{tab3}
\end{table}

As illustrated in Fig.~\ref{fig5}, the proposed RAIL algorithm, represented by the red line, consistently achieves lower localization error compared to the Min-Max and RSSI-based DV-hop algorithms. This reflects the superior accuracy of RAIL. Furthermore, the error distribution of RAIL is more consistent, indicating better stability of the algorithm. The data presented in TABLE~\ref{tab3} corroborate this observation.

When the number of nodes is relatively small, especially when the number of nodes is 100, all three algorithms occasionally produce significant localization errors. This occurs because sparse node deployment leads to areas in the network lacking node coverage. In such cases, the estimated distance from the anchor node to the unknown node in the Min-Max and DV-hop algorithms often significantly exceeds the true distance. For example, the maximum observed error for Min-Max is 44.7166 meters. Its largest error reaches 36.8491 meters. This is primarily because the shortest path between the anchor and the unknown node becomes highly tortuous due to node sparsity, resulting in angle estimation deviation. However, the bounding box constraint in RAIL helps mitigate this error, offering a level of robustness that the other methods lack. As node density increases, this issue diminishes. Particularly at 500 nodes, large errors across all three algorithms occur rarely.

The average error for the RAIL algorithms decreases by 47.16\% as node count rises. In contrast, the Min-Max algorithm shows little improvement. For RAIL, more nodes around each anchor improve angle inference, while simultaneously refining the bounding box, which further reduces localization error. However, since Min-Max does not rely on the ranging method, it is more random and strongly depends on the number and relative position relationship of the anchor node and the unknown node. When the anchor node is on the same side of the unknown node, a large error will occur. Therefore, the increase in the number of nodes does not improve the accuracy of this method much.

In terms of stability, the standard deviation analysis underscores the performance advantages of RAIL. In terms of standard deviation error, the proposed RAIL algorithm achieves reductions of 56.5\% compared to the Min-Max algorithm and 25.5\% compared to the RSSI-based DV-hop algorithm. Notably, both RAIL and DV-hop show reduced deviation as node density increases, whereas Min-Max remains no significant changes. A smaller standard deviation signifies a more reliable and robust localization algorithm.

In summary, the RAIL algorithm outperforms Min-Max and RSSI-based DV-hop in both accuracy and stability. Specifically, at 100 nodes, RAIL’s average error is 59.2\% and 47.2\% lower than Min-Max and DV-hop, respectively. At 200 nodes, the reductions are 76.9\% and 56.7\%, and at 500 nodes, 81.2\% and 60.2\%. These results demonstrate that RAIL’s advantages become more pronounced with increasing network density, making it a highly accurate and reliable localization solution for wireless sensor networks.

\section{Conclusion}
Accurate and fast localization of nodes in WSNs is a task of significant importance but inherent complexity. To address this challenge, this paper proposes a general-purpose localization algorithm based on RSSI. By inferring the angle between unknown nodes and anchor nodes, the proposed method enables effective position estimation without additional hardware requirements. The algorithm demonstrates strong practicality and compatibility with AI applications, especially in scenarios requiring data generation or model pretraining. Under varying node densities and with only three anchor nodes, the proposed RAIL algorithm achieves an average localization error of 4.0478 meters. Compared with the Min-Max algorithm and RSSI-based DV-hop, RAIL achieves an overall performance improvement of 72.4\%. The high accuracy and stability of the RAIL algorithm indicate its strong applicability in real-world deployments, particularly in dynamic and resource-constrained environments.

\end{document}